\documentclass{WileyMSP-template}

\begin{document}

\pagestyle{fancy}
\rhead{\includegraphics[width=2.5cm]{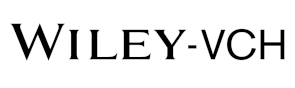}}

\title{Exploring $\gamma$-In$_2$(Se$_{1-x}$Te$_x$)$_3$ alloys as photovoltaic materials}

\maketitle


\author{Wei Li}
\author{Xuefen Cai}
\author{Nicholas Valdes}
\author{Tianshi Wang}
\author{William Shafarman}
\author{Su-Huai Wei}
\author{Anderson Janotti}



\begin{affiliations}

Wei Li, Tianshi Wang\\
Department of Materials Science \& Engineering, University of Delaware, Newark, DE 19716, USA\\
Xuefen Cai\\
Department of Materials Science \& Engineering, University of Delaware, Newark, DE 19716, USA\\
Beijing Computational Science Research Center, Beijing 100193, China\\

Nicholas Valdes, William Shafarman\\
Department of Materials Science \& Engineering and Institute of Energy Conversion, University of Delaware, Newark, DE 19716, USA\\

Su-Huai Wei\\
Beijing Computational Science Research Center, Beijing 100193, China\\

Anderson Janotti\\
Department of Materials Science \& Engineering, University of Delaware, Newark, DE 19716, USA\\
janotti@udel.edu

\end{affiliations}



\begin{abstract}

In$_2$Se$_3$ in the three-dimensional (3D) hexagonal crystal structure with space group $P6_1$ ($\gamma$-In$_2$Se$_3$) has a direct band gap of $\sim$1.8 eV and high absorption coefficient, making it a promising semiconductor material for optoelectronics. Incorporating Te allows for tuning the band gap, adding flexibility to device design and extending the application range. Here we report the growth and characterization of $\gamma$-In$_2$Se$_3$ thin films, and results of hybrid density functional theory calculations to assess the electronic and optical properties of $\gamma$-In$_2$Se$_3$ and $\gamma$-In$_2$(Se$_{1-x}$Te$_x$)$_3$ alloys. The calculated band gap of 1.84 eV for $\gamma$-In$_2$Se$_3$ is in good agreement with data from the absorption spectrum, and the absorption coefficient is found to be as high as that of direct band gap conventional III-V and II-VI semiconductors.  Incorporation of Te in the form of $\gamma$-In$_2$(Se$_{1-x}$Te$_x$)$_3$ alloys is an effective way to tune the band gap from 1.84 eV down to 1.23 eV, thus covering the optimal band gap range for solar cells.
We  also discuss band-gap bowing and mixing enthalpies, aiming at adding $\gamma$-In$_2$Se$_3$ and $\gamma$-In$_2$(Se$_{1-x}$Te$_x$)$_3$ alloys to the available toolbox of materials for solar cells and other optoelectronic devices.

\end{abstract}

\section{Introduction}

\begin{figure}[ht!]
\begin{center}
\includegraphics[width=4.0in]{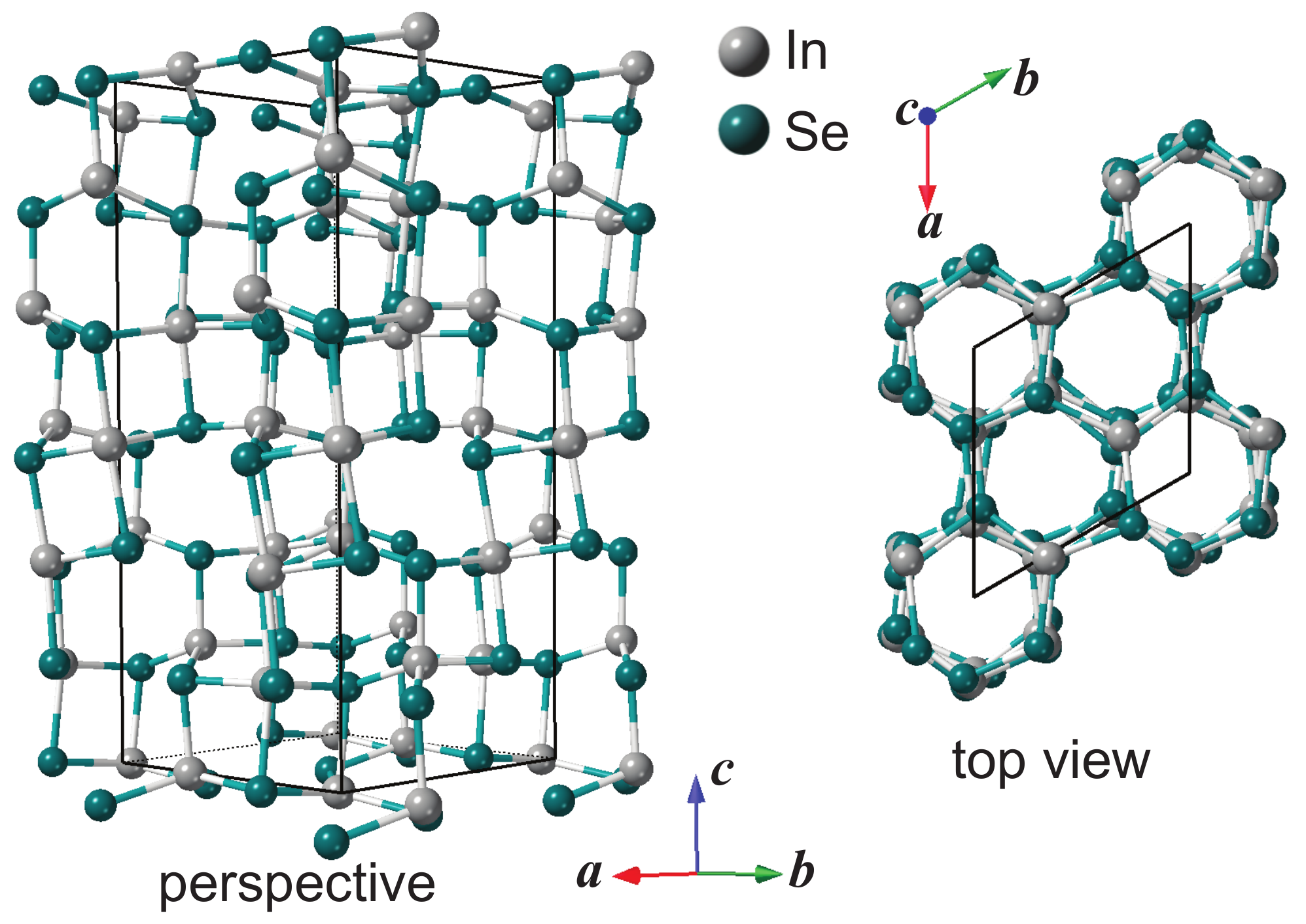}
\end{center}
\caption{ Crystal structure of $\gamma$-In$_2$Se$_3$, showing a perspective and a top view of the defective wurtzite crystal structure with space group $P6_1$.
}
\label{fig1}
\end{figure}

In$_2$Se$_3$ crystallizes in a defect wurtzite-like structure ($\gamma$-In$_2$Se$_3$, Fig.~\ref{fig1}), which appears to be the most stable phase of In$_2$Se$_3$ at room temperature. It has a direct band gap of $\sim$1.8 eV, and displays a high absorption coefficient in the visible-light range \cite{Zheng2016,DeGroot2001,Julien1990,Chaiken2003,Ohtake1997,Nakayama2005,Ye1998,Lyu2010}. In$_2$Se$_3$ thin film has been used as a precursor layer in co-evaporation of CuInSe$_{2}$-based solar cells \cite{Contreras1994}, as well as a buffer layer to replace CdS in Mo/CIS/In$_2$Se$_3$/ZnO device structures, showing higher open-circuit voltage than the Mo/CIS/CdS/ZnO structure \cite{Gordillo2003}. The band gap of In$_2$Se$_3$ can be lowered by adding Te \cite{Emziane2001} in the form of In$_2$(Se,Te)$_3$ alloys, covering the spectrum region suitable for solar cell applications. Besides this basic knowledge, details of the electronic and optical propeties of In$_2$Se$_3$, In$_2$Te$_3$, and In$_2$(Se,Te)$_3$ alloys have remained largely unexplored.

Here we report a combination of experimental and computational studies on
$\gamma$-In$_2$Se$_3$ and $\gamma$-In$_2$(Se$_{1-x}$Te$_x$)$_3$ alloys, focusing on their basic structural, electronic, and optical properties, with the aim of developing these materials for electronic and optoelectronic applications. First, we discuss the growth and characterization of In$_2$Se$_3$ thin films, followed by a discussion of hybrid density functional calculations of basic properties of In$_2$Se$_3$ and a comparison with experimental data. We then report on the absorption coefficients and band alignment between $\gamma$-In$_2$Se$_3$ and $\gamma$-In$_2$Te$_3$, and with other more conventional chalcogenide semiconductors, such as CdTe and CuInSe$_2$. Finally, we discuss the stability and electronic properties of $\gamma$-In$_2$(Se$_{1-x}$Te$_x$)$_3$ alloys, showing that the band gap can be adjusted from 1.84 eV down to 1.23 by changing the concentration of Te, passing through the optimum range of direct band gaps for a single-junction solar cell. We then conclude with a discussion on how this materials system offers new opportunities for innovative architectures and device design.

\section{Experimental and Computational Approaches}

In$_2$Se$_3$ was grown on soda lime glass substrates by thermal co-evaporation of In and Se. The growth lasted 30 min and the substrate temperature was 350 $^{\circ}$C to form films with thickness of 1$\mu m$. Composition was verified by x-ray fluorescence. Spectrophotometry was performed to measure transmission and reflection in the wavelength range 300-2000 nm, and the absorption coefficient was determined from the reflectance and transmission \cite{Pankove}.

The calculations were based on the density functional theory \cite{Kohn1964,Kohn1965} as implemented in the \textsc{VASP} code\cite{Kresse1993a,Kresse1993b}. The interactions between the valence electrons and the ionic cores are described using projector augmented wave (PAW) potentials \cite{Blochl1994,Kresse1999}. We used a kinetic energy cutoff of 320 eV for the plane wave basis set.  Structure optimization was performed using the Perdew-Burke-Ernzerhof functional revised for solids (PBEsol) \cite{Perdew} on a $\Gamma$-centered 6$\times$6$\times$2 mesh for the 30-atom primitive cells of $\gamma$-In$_2$Se$_3$ and $\gamma$-In$_2$Te$_3$, and equivalent $k$-point mesh for supercells for the $\gamma$-In$_2$(Se$_{1-x}$Te$_x$)$_3$ alloys. Once the ionic positions were determined using the PBEsol functional, we calculated the electronic structure and dielectric functions using the HSE06 functional \cite{Heyd2003,Heyd2006}, including the effects of spin-orbit coupling. Note that the band gaps consistently calculated in HSE06 using optimized lattice parameters also in HSE06 are only slightly higher in energy by less than 0.1 eV. Contributions from excitons and phonon-assisted optical transitions to the absorption coefficients, expected to be small, were not included in the present work. To obtain a good description of optical band gap threshold, we use the tetrahedral method with a small complex shift of 1 meV in the Kramers-Kronig transformation \cite{Toll1956}.

For simulating the In$_2$(Se$_{1-x}$Te$_x$)$_3$ random alloys we employed special quasi-random structures (SQS) \cite{Zunger1990, Wei1990} obtained using the Alloy Theoretic Automated Toolkit (ATAT) \cite{VanDeWalle2013}. This approach is based on a Monte Carlo simulated annealing loop with an objective function that seeks to perfectly match the maximum number of atomic correlation functions of the random alloys\cite{VanDeWalle2013}. Here we use supercells with 90 atoms for representing In$_2$(Se$_{1-x}$Te$_x$)$_3$ alloys.

\begin{figure}[ht]
\begin{center}
\includegraphics[width=4.4 in]{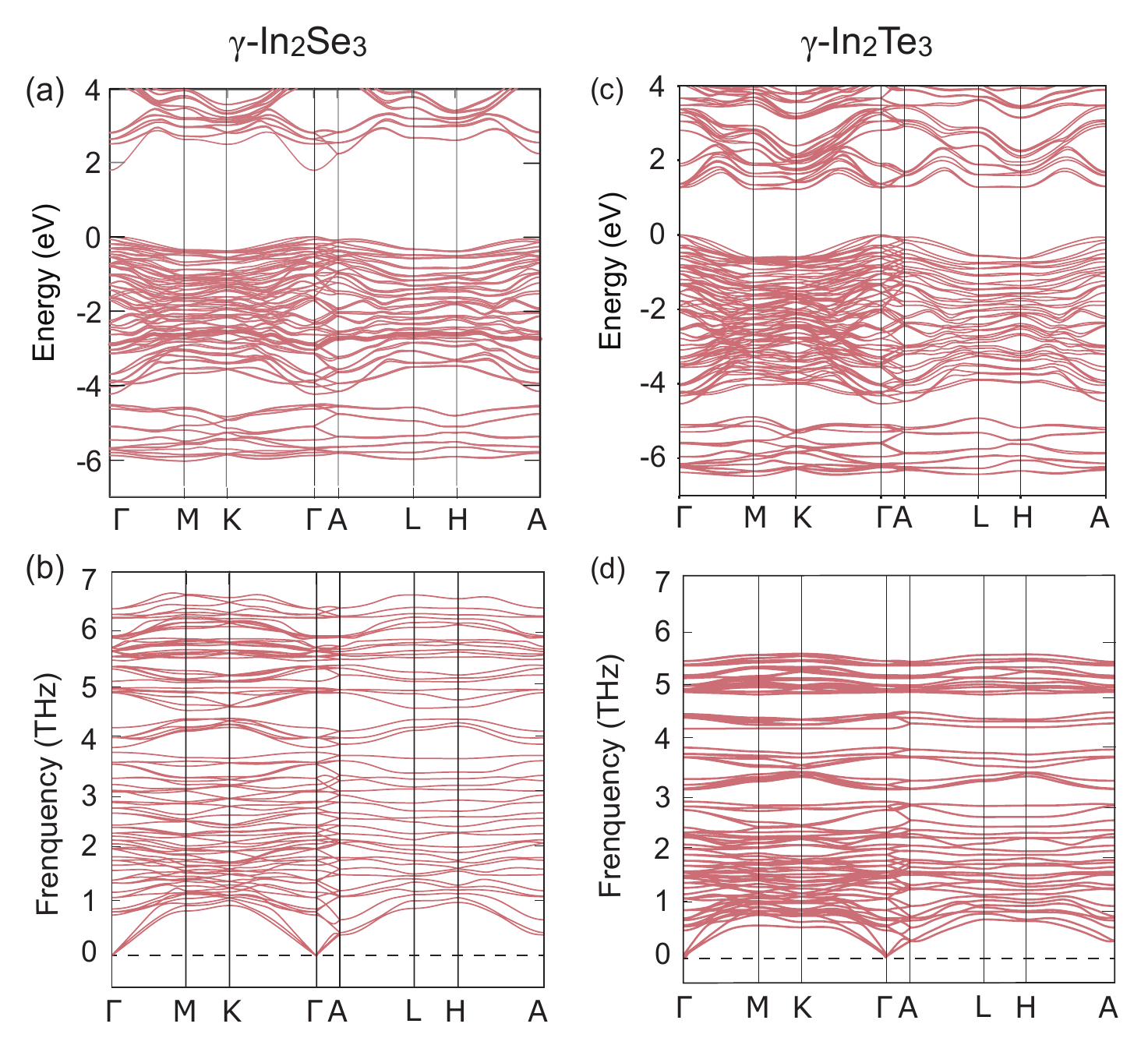}
\end{center}
\caption{
Electronic band structure and phonon dispersion of $\gamma$-In$_2$Se$_3$ and $\gamma$-In$_2$Te$_3$. Calculated electronic structure of (a) $\gamma$-In$_2$Se$_3$ showing a direct band gap at $\Gamma$, and (c) of $\gamma$-In$_2$Te$_3$ showing the direct band gap at $\Gamma$ very close in energy to the indirect band gap with VBM at $\Gamma$ and CBM at ${\rm K}$.
Calculated phonon dispersion of (b) $\gamma$-In$_2$Se$_3$ and (d) $\gamma$-In$_2$Te$_3$ along high-symmetry directions; the absence of negative phonon frequencies indicates the structural stability of the of the two compounds in the $\gamma$ phase.
}
\label{fig2}
\end{figure}

\section{Results and discussion}

\subsection{Structural and electronic properties of $\gamma$-In$_2$Se$_3$ and $\gamma$-In$_2$Te$_3$}

The defective wurtzite crystal structure of $\gamma$-In$_2$Se$_3$ (and $\gamma$-In$_2$Te$_3$) with space group $P6_1$ is shown in Fig.~\ref{fig1}. This crystal structure is built up of In(1)Se$_5$ trigonal bipyramids and In(2)Se$_4$ tetrahedra \cite{Pfitzner1996}, which are connected by common corners and edges resulting in a distorted wurtzite-like structure, i.e., a wurtzite structure missing one third of cation sites. The vacant sites are orderly arranged forming a screw along the $c$ axis (vacancy ordered in screw form, or VOSF phase) \cite{Peng2009,Ye1998,Han2014}. The calculated equilibrium lattice parameters for $\gamma$-In$_2$Se$_3$ are $a=7.179$ {\AA} and $c=19.412$ {\AA}, in good agreement with experimental data $a=7.129$ {\AA} and $c=19.381$ {\AA} \cite{Pfitzner1996}. For $\gamma$-In$_2$Te$_3$, the calculated lattice constants are $a=7.632$ {\AA} and $c=$20.980 {\AA}.

\begin{figure}[ht]
\begin{center}
\includegraphics[width=3.2in]{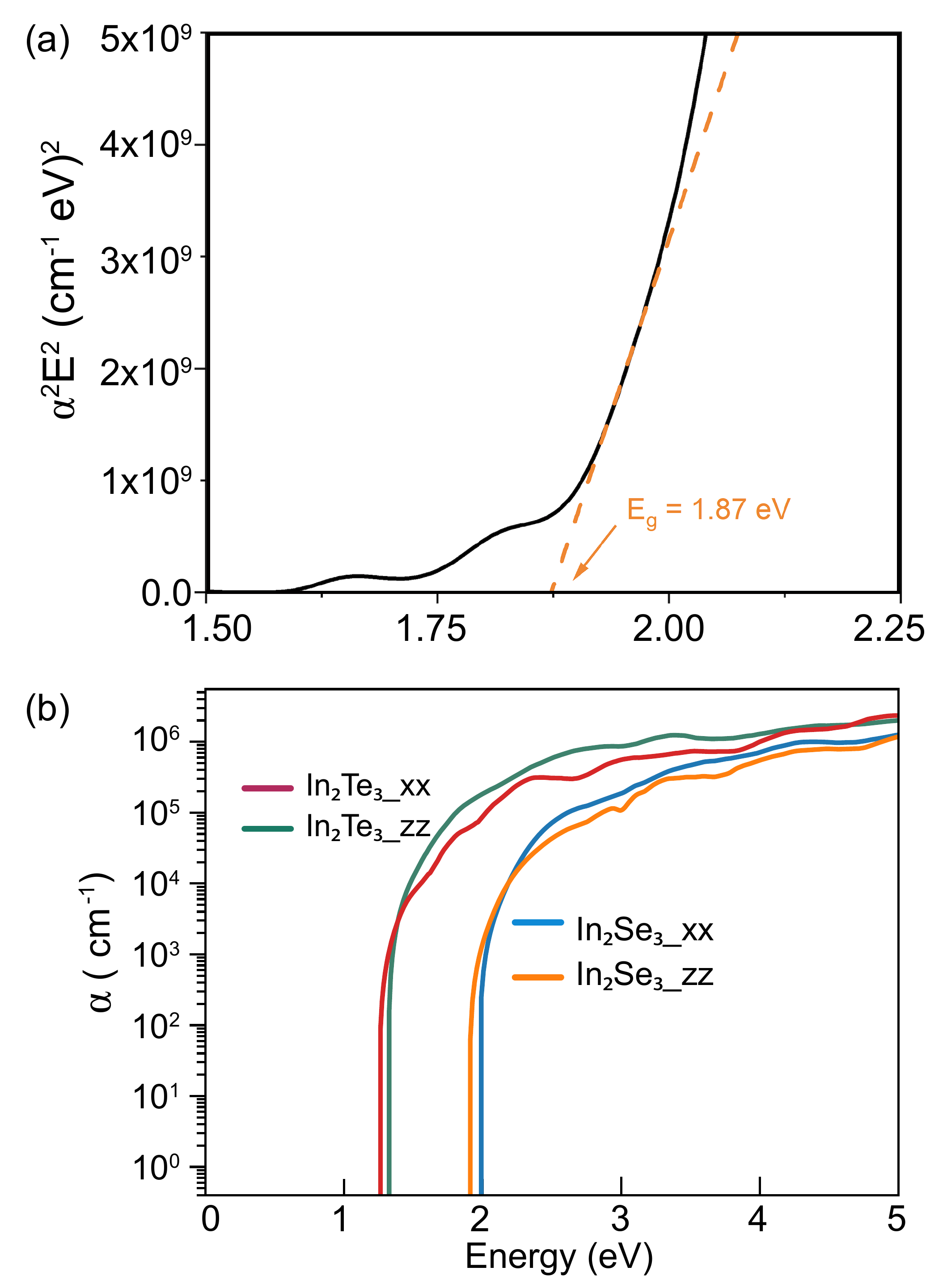}
\end{center}
\caption{Optical properties of $\gamma$-In$_2$Se$_3$ and $\gamma$-In$_2$Te$_3$: (a)
Tauc plot of the optical absorption derived from the transmission and reflection spectra of the In$_2$Se$_3$ thin film, indicating a gap of 1.87 eV, and (b) calculated absoption coefficient ($\alpha$) of bulk $\gamma$-In$_2$Se$_3$ and $\gamma$-In$_2$Te$_3$ in log scale.}
\label{fig3}
\end{figure}

The calculated electronic structure for $\gamma$-In$_2$Se$_3$ and $\gamma$-In$_2$Te$_3$ are shown in Figs.~\ref{fig2}(a) and (c). In the case of $\gamma$-In$_2$Se$_3$, we find a direct band gap of 1.84 eV at the $\Gamma$ point, in good agreement with the onset of optical absorption shown in Fig.~\ref{fig3}(a) and previously reported values of 1.8-1.9 eV \cite{DeGroot2001,Julien1990,Chaiken2003}. Previous DFT-GGA calculations for $\gamma$-In$_2$Se$_3$ showed a band gap of $\sim$1.0 eV \cite{Cui2017}. However, GGA functionals are known to underestimate band gaps. We find that the lowest conduction band originates from In $5s$ orbitals, while the highest valence band originates mostly from Se $4p$ orbitals. The calculated effective electron mass values are 0.136 m$_{e}$ and 0.148 m$_{e}$, respectively, along the in-plane and out-of-plane directions.
In the case of $\gamma$-In$_2$Te$_3$, we find an indirect band gap of 1.23 eV, where the conduction-band minimum (CBM) at K is only slightly lower than at the $\Gamma$ point by 0.04 eV. Note that our HSE06 calculations using the HSE06 optimized lattice parameters give a direct band gap at $\Gamma$  of 1.34 eV, with the indirect gap $\Gamma$-K slightly higher in energy.  Similar to In$_{2}$Se$_{3}$, the lowest energy conduction band is derived from In $5s$ orbitals and the valence band maximum (VBM) state at $\Gamma$ is mainly derived from Te $5p$ orbitals.

The calculated phonon dispersions of $\gamma$-In$_2$Se$_3$ and $\gamma$-In$_2$Te$_3$, shown in Fig.~\ref{fig2}(b) and (d), reveals no trace of imaginary frequency, indicating that our optimized structure for $\gamma$-In$_2$Se$_3$ represents a minimum-energy structure, consistent with experimental results indicating that this phase is thermodynamically stable at low temperatures \cite{Ye1998,Amory2003,Lyu2010}.

For the optical properties, we show the absorption spectra of $\gamma$-In$_2$Se$_3$ and $\gamma$-In$_2$Te$_3$ in Fig.~\ref{fig3}. A Tauc plot of the optical absorption absorption of the In$_2$Se$_3$ thin film, obtained from measured transmission and reflection spectra, is shown in Fig.~\ref{fig3}(a), with the obtained direct band gap of 1.87 indicated by the arrow.  For comparison, the calculated absorption coefficient $\alpha$ of bulk $\gamma$-In$_2$Se$_3$ and $\gamma$-In$_2$Te$_3$ are shown in Fig.~\ref{fig3}(b). We find a good agreement between the calculated band gap of 1.84 eV and the value of 1.87 eV obtained from the onset in the Tauc plot \cite{Sreekumar2008,Emziane1997,Groot2001}.

Due to the hexagonal crystal structure, the absorption coefficient is anisotropic, depending on the light polarization being parallel or perpendicular to the $c_{hex}$ axis.
The absorption coefficient of $\gamma$-In$_2$Se$_3$ and $\gamma$-In$_2$Te$_3$, derived from the real and imaginary parts of the dielectric function, rapidly increases for photon energies above the band gap, and are comparable to those of conventional III-V and II-VI semiconductors. For
$\gamma$-In$_2$Te$_3$, we note that the calculated onset at 1.27 eV in the absorption coefficient  occurs at photon energies that are slightly higher than the calculated fundamental indirect band gap.

\begin{figure}[ht!]
\begin{center}
\includegraphics[width=4.0in]{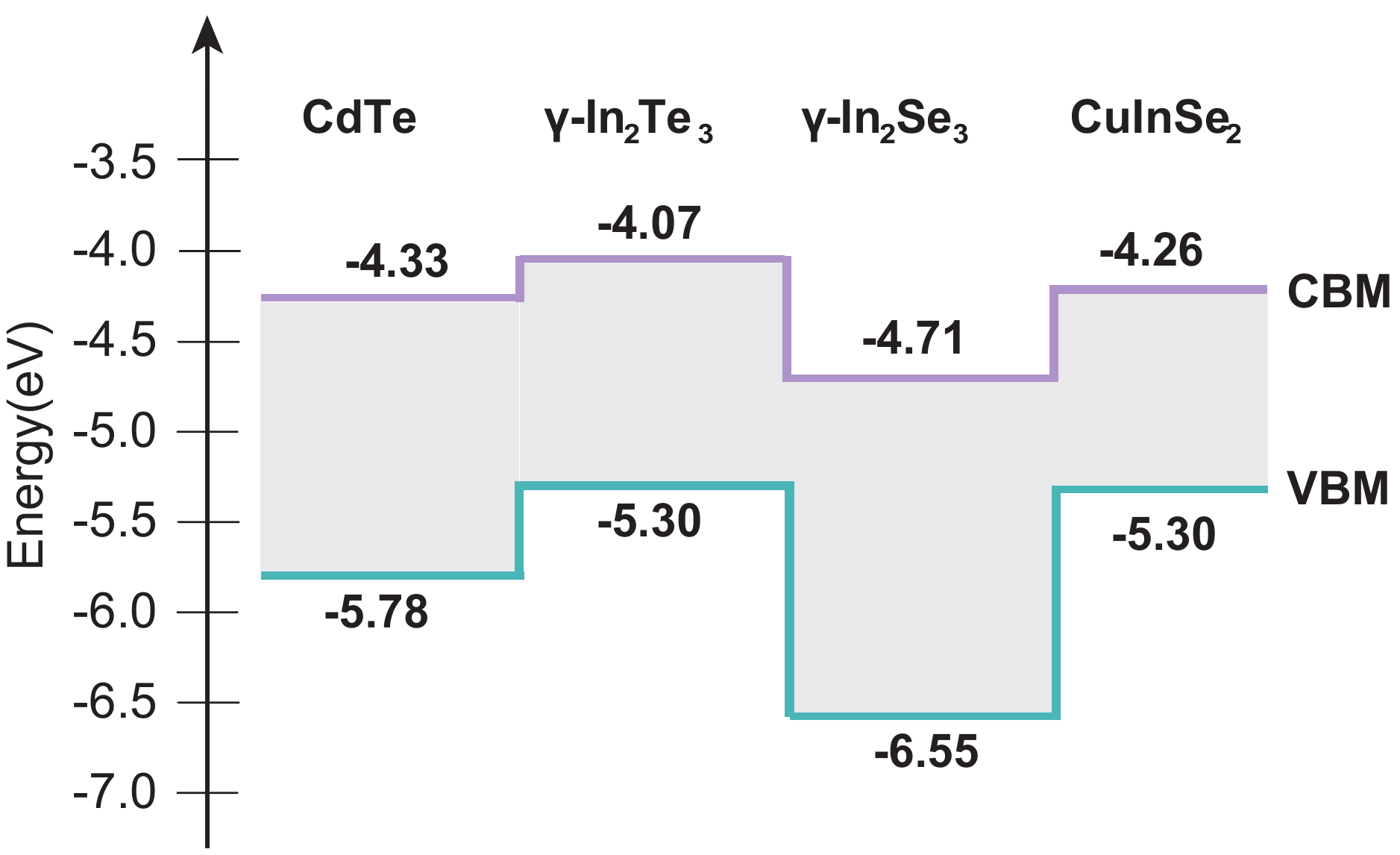}
\end{center}
\caption{Band alignments and band edge positions with respect to vacuum for $\gamma$-In$_2$Se$_3$ and $\gamma$-In$_2$Te$_3$, compared to CuInSe$_{2}$ (from Ref.~\cite{Wei1993}) and CdTe (from Ref.~\cite{Oba2014}), which are two important materials for thin-film solar cells. }
\label{fig4}
\end{figure}

For semiconductors, besides the band gap, it is essential to know the position of the valence and conduction band edges in an absolute
energy scale, and the band alignment to other common semiconductors for designing contacts and heterojunctions.
In this context, we calculate the band edge energies of $\gamma$-In$_2$Se$_3$ and $\gamma$-In$_2$Te$_3$ with respect to the vacuum level
using the following procedure: First we built slabs for $\beta$-In$_2$Se$_3$ and $\beta$-In$_2$Te$_3$ constraining the volume per formula unit to be the same as that of $\gamma$-In$_2$Se$_3$ and $\gamma$-In$_2$Te$_3$, respectively. The $\beta$ crystal structure is the same as that of Bi${_2}$Se${_3}$, which is a hexagonal two-dimensional layered structure formed of quintuple layers \cite{Li2018}, with a natural cleavage plane perpendicular to the $c$-axis. Since the volume per formula unit of the constructed $\beta$ phase is constrained to be the same of that of the $\gamma$ phase, they have the same average electrostatic potential.  From the slab of this constrained $\beta$ phase, we obtain the average electrostatic potential in a bulk region with respect to the vacuum level. We then combine the top of the valence band of the $\gamma$ phase with respect to the average electrostatic potential in a bulk calculation, and the position of the average electrostatic potential of the $\beta$ phase with respect to vacuum to obtain the position of the valence band of the $\gamma$ phase with respect to vacuum.  The results for $\gamma$-In$_2$Se$_3$ and $\gamma$-In$_2$Te$_3$ are shown in Fig.~\ref{fig4}, along with the values for CdTe and CuInSe$_{2}$ from the literature \cite{Wei1993,Oba2014}.  First, we see that the valence-band maximum of $\gamma$-In$_2$Te$_3$ is 1.25 higher than that of $\gamma$-In$_2$Se$_3$, which is attributed to the higher energy of Te $5p$ than the Se $4p$. This difference is larger than those between CdTe and CdSe and between ZnTe and ZnSe \cite{Oba2014,Yang2019}.  This can be explained by the fact that in these II-VI compounds there is a strong $p$-$d$ coupling \cite{Wei2000,Yang2019} that pushes up the valence-band maximum, and it is stronger in the selenides than in the tellurides, reducing the valence band offset between them.

As also shown in Fig.~\ref{fig4}, we find that the valence-band maximum of $\gamma$-In$_2$Te$_3$ is higher than that of CdTe, and this is because Te is three-fold coordinated in $\gamma$-In$_2$Te$_3$, resulting in a non-bonding or lone-pair character of the valence-band maximum that is higher in energy than the valence-band-maximum of CdTe with a mainly Te $p$-Cd $p$ bonding character.
Note that CdTe is a material that is used in thin film solar cells, and one of the current limitations of CdTe-based solar cells is the efficiency of $p$-type doping. Having a valence-band maximum higher than that of CdTe, we expect that it would be easier to dope $\gamma$-In$_2$Te$_3$ $p$-type, thus making this material itself a promising material for thin film solar cells.

\begin{figure}[ht]
\begin{center}
\includegraphics[width=4.8 in]{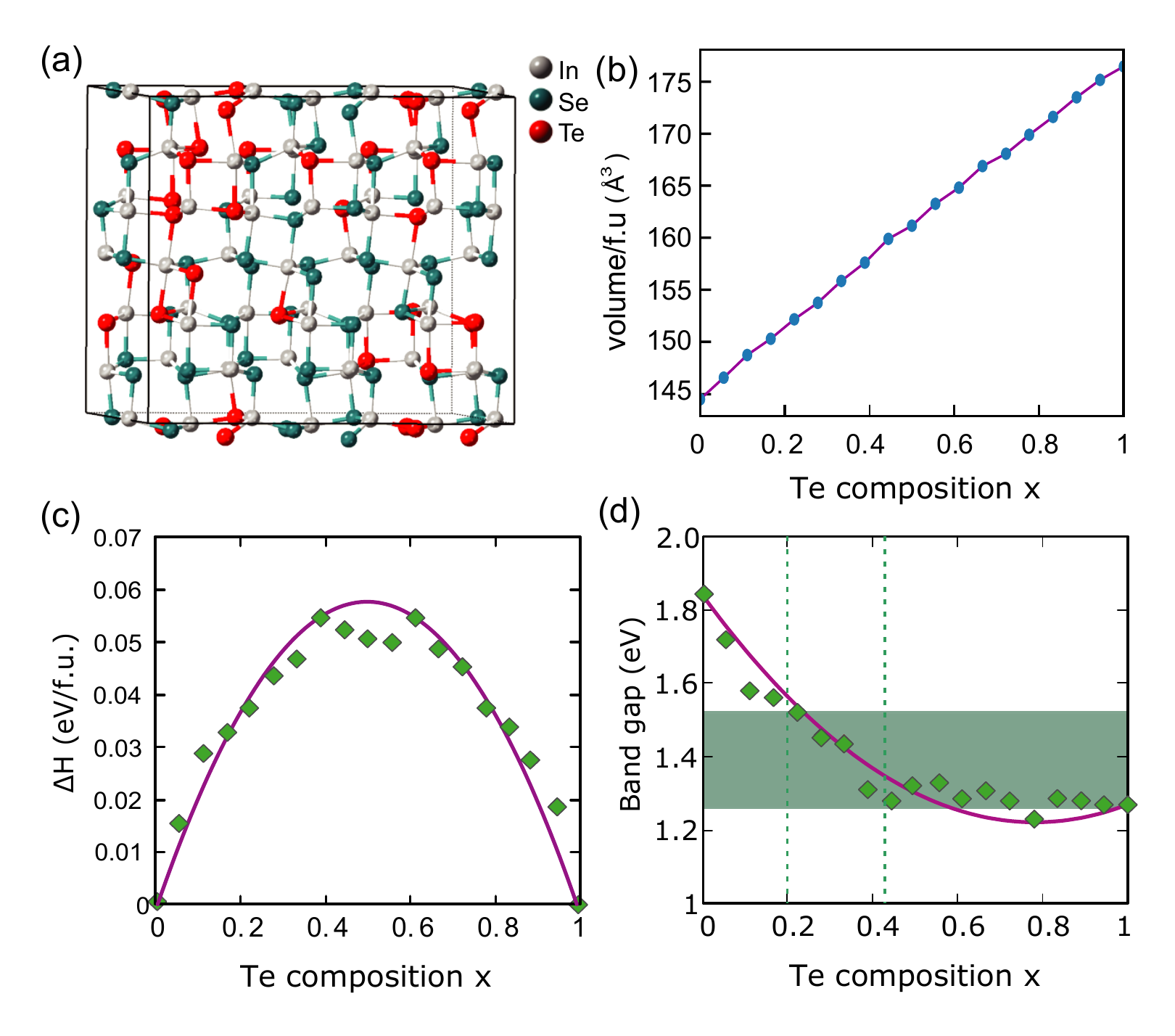}
\end{center}
\caption{ Structure, volume, mixing enthalpy, and band gap of $\gamma$-In$_{2}$(Se$_{1-x}$Te$_{x}$)$_{3}$ alloys. The structure in (a) represents a SQS structure for $x=0.3$; the volume per formula unit of the alloys in (b) follows Vergard's law; The mixing enthalpy per formula unit as a function of Te concentration $x$ in (c) is relatively low and comparable to other semiconductor alloy systems for which mixing over the whole concentration range has been observed. (d) Band gap of $\gamma$-In$_{2}$(Se$_{1-x}$Te$_{x}$)$_{3}$ alloys as a function of Te concentration $x$.
}
\label{fig5}
\end{figure}

\subsection{Structure, stability, and band gap of $\gamma$-In$_2$(Se$_{1-x}$Te$_x$)$_3$ alloys}

Having established the structural, electronic, and optical properties of bulk $\gamma$-In$_2$Se$_3$ and $\gamma$-In$_2$Te$_3$, we now discuss the basic properties of $\gamma$-In$_{2}$(Se$_{1-x}$Te$_{x}$)$_{3}$ alloys. We simulate these random alloys using SQS structures in a supercell of 90 atoms, varying the Se/Te concentrations in steps of $\Delta x$=0.056, from $x=0$ ($\gamma$-In$_2$Se$_3$) to $x=1$ ($\gamma$-In$_2$Te$_3$). One example of such SQS structure for $x=0.3$ is shown in Fig.~\ref{fig5}(a).  In  Fig.\ref{fig5}(b) we show the evolution of the volume per formula unit of $\gamma$-In$_{2}$(Se$_{1-x}$Te$_{x}$)$_{3}$ alloys as a function of Te composition $x$, indicating that they closely follow Vegard's law\cite{Vegard1921}.

The formability of $\gamma$-In$_{2}$(Se$_{1-x}$Te$_{x}$)$_{3}$ alloys is determined by calculating the mixing enthalpy, defined by:
\begin{equation}
\Delta H_f[x] = E[x]-x E[\gamma{\rm-In}_2{\rm Se}_3]- (1-x) E[\gamma{\rm-In}_2{\rm Te}_3],
\label{enthalpy}
\end{equation}
where $E[x]$ is the total energy of the $\gamma$-In$_{2}$(Se$_{1-x}$Te$_{x}$)$_{3}$ SQS supercell for a given $x$, $E[\gamma{\rm-In}_2{\rm Se}_3]$ is the total energy of bulk $\gamma$-In$_2$Se$_3$ and $E[\gamma{\rm-In}_2{\rm Te}_3]$ is the total energy of bulk $\gamma$-In$_2$Te$_3$ using the same supercell size and $k$-mesh of the SQS structure. The results are shown in Fig.~\ref{fig5}(c).  We find that the mixing enthalpies are all positive and comparable to those of III-V-based alloys \cite{Ferreira1989}, where uniform mixing have been observed for all alloy concentrations. The positive sign of $\Delta H_f$ indicates that the ground state of these alloys at $T=0$ corresponds to phase separation into the binary constituents. However, since the formation enthalpies are relatively low, these alloys are likely to be stabilized through entropy at finite temperatures for the entire range of Te compositions.

The calculated band gap of $\gamma$-In$_{2}$(Se$_{1-x}$Te$_{x}$)$_{3}$ alloys as a function of composition $x$ is shown in Fig.~\ref{fig5}(d). Note that these results also include the effects of spin-orbit coupling. Here we only discuss the direct band gap at $\Gamma$ in the alloys, despite the conduction-band minimum at K and at $\Gamma$ being very close in energy in $\gamma$-In$_2$Te$_3$. The calculated bowing parameter $b$, defined as \cite{Bernard1987},
 \begin{equation}
E_{g}[x]= (1-x)E_{g}[\gamma{\rm-In}_2{\rm Se}_3]+x E_{g}[\gamma{\rm-In}_2{\rm Te}_3]- b x(1-x),
\label{eq2}
\end{equation}
is 1.0 eV, which is close to other chalcogenide alloys such as (Cd,Se)Te (0.83 eV) and (Zn,Se)Te (1.23 eV) \cite{Wei1995,Tit_2010,Wei2000}.

The band gap of $\gamma$-In$_{2}$(Se$_{1-x}$Te$_{x}$)$_{3}$ rapidly decreases with Te composition for small $x$, reaching around 1.3 eV for $x=0.4$. For larger Te compositions, $x>0.4$, the gap remains within 0.1 eV of the gap of  $\gamma$-In$_2$Te$_3$ and has a minimum band gap of 1.2 eV at $x=0.8$. We attribute this behavior to the large difference between the energy of the Se/Te valence $p$ orbitals, i.e., adding small amounts of Te to $\gamma$-In$_2$Se$_3$ leads to an isovalent Te-related band that is expected to be much higher in energy than the host valence band. Such a high VBM of the alloy not only reduces the band gap, but also facilitates $p$-type doping as in pure $\gamma$-In$_2$Te$_3$. Conversely, adding small amounts of Se to $\gamma$-In$_2$Te$_3$ would lead to an isovalent Se-related band below the VBM of the host, causing only small changes in the band gap.  Note that the band gap of $\gamma$-In$_{2}$(Se$_{1-x}$Te$_{x}$)$_{3}$ alloys can be tailored to be in the range of 1.2-1.5 eV when the Te concentration is between 20\% and 100\%, which is recognized as the optimum range of band gaps for single junction solar cells.

\section{Conclusion}

In summary, we investigated the basic structural, electronic, and optical properties of $\gamma$-In$_2$Se$_3$, both experimentally and computationally. We predicted the properties of $\gamma$-In$_2$Te$_3$,
and $\gamma$-In$_{2}$(Se$_{1-x}$Te$_{x}$)$_{3}$ alloys, aiming at developing these materials for photovoltaic and other optoelectronic applications. $\gamma$-In$_2$Se$_3$ has a direct band gap of 1.84 eV, in good agreement with the onset in the absorption spectrum, whereas $\gamma$-In$_2$Te$_3$ has an indirect band gap of 1.23 eV, with the direct band gap at $\Gamma$ being only 0.04 eV higher. The absorption coefficients of $\gamma$-In$_2$Se$_3$ and $\gamma$-In$_2$Te$_3$ are very high in the visible light range, reaching 10$^{5}$ cm$^{-1}$ at $\sim$1 eV above the band gap energy. The valence band of $\gamma$-In$_2$Te$_3$ is predicted to be higher than that of CdTe, indicating that this material can be easily doped $p$-type. $\gamma$-In$_{2}$(Se$_{1-x}$Te$_{x}$)$_{3}$ alloys are predicted to easily form, with a maximum mixing enthalpy of 52 meV/f.u., similar to other semiconductor alloys for which mixing over the whole concentration range has been observed. The band gap of $\gamma$-In$_{2}$(Se$_{1-x}$Te$_{x}$)$_{3}$ alloys rapidly decreases with increasing Te composition, falls in the range of 1.2-1.5 eV for Te concentrations higher than $\sim$20\%, which is optimal for single junction solar cells. With such high Te concentration, the VBM is also high enough so the alloy could be easily doped p-type. Our work thus indicates that the $\gamma$-In$_{2}$(Se$_{1-x}$Te$_{x}$)$_{3}$ alloy system is promising for photovoltaic applications.

\medskip
\textbf{Acknowledgements} \par 
This work was supported by the National Science Foundation Faculty Early Career Development Program DMR-1652994. This research was also supported by the eXtreme Science and Engineering Discovery Environment (XSEDE) facility, National Science Foundation grant number ACI-1053575, and the Information Technologies (IT) resources at the University of Delaware. NV acknowledges support from National Science Foundation under award number 1507351.
XC and SHW acknowledge the support from the Beijing Science and Technology Committee (No. Z181100005118003),
National Nature Science Foundation of China (No. 51672023; 11634003; U1930402) and computational support from the Beijing Computational Science Research Center (CSRC), and support from the China Scholarship Council (No. 201904890014).

\medskip

%
\bibliographystyle{MSP}


\begin{figure}
\textbf{Table of Contents}\\
\medskip
  \includegraphics[width=3in]{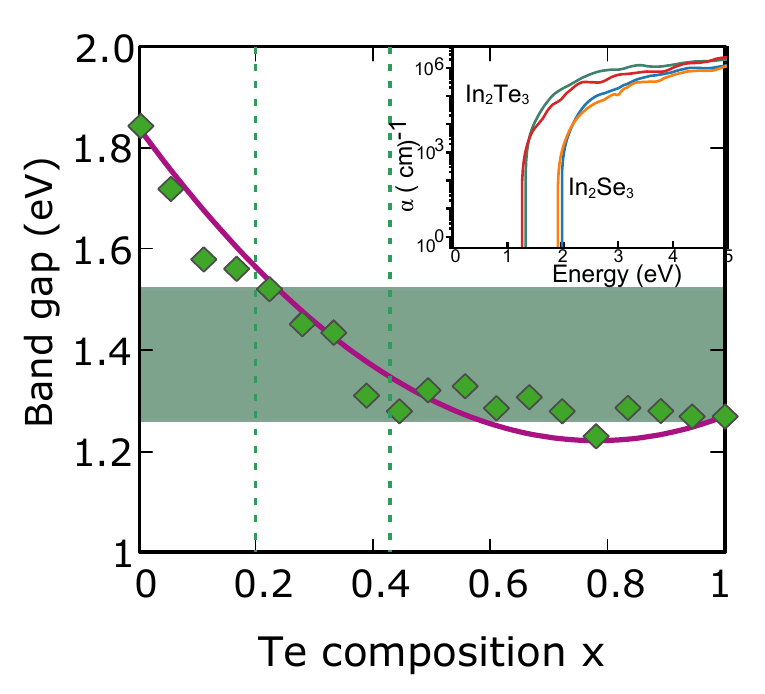}
  \medskip
\end{figure}

\end{document}